\documentstyle[12pt,aps]{revtex}

\include{psfig}

\input epsf

\begin{document}

\title{
\hfill\parbox[t]{2in}{\rm\small\baselineskip 14pt
{JLAB-THY-01-05}\vfill~}
\vskip 2cm
Evidence Against Instanton Dominance\\
of Topological Charge Fluctuations in QCD \\ ~~\\
}
\author{Ivan Horv\'ath}
\address{Department of Physics and Astronomy, University of Kentucky,
Lexington, KY 40506}

\author{Nathan Isgur}
\address{Jefferson Lab, 12000 Jefferson Avenue,
Newport News, VA  23606}

\author{John McCune and H. B. Thacker}
\address{Department of Physics, University of Virginia,
Charlottesville, VA 22901}

\maketitle

\vskip 0.2in

\begin{abstract}

The low-lying eigenmodes of the Dirac operator associated with typical
gauge field configurations in QCD encode, among other low-energy properties, 
the physics behind the solution to the $U_A(1)$ problem
({\it i.e.} the origin of the $\eta'$ mass), the nature of spontaneous chiral symmetry breaking,
and the physics of string-breaking, quark-antiquark pair production, and the OZI rule. 
Moreover, the space-time chiral structure of these eigenmodes reflects the 
space-time topological structure of the underlying gauge field. We present 
evidence from lattice QCD on the {\it local chiral structure} of low Dirac eigenmodes
leading to the conclusion that topological 
charge fluctuations of the QCD vacuum are not instanton-dominated. The 
result supports Witten's arguments that topological charge is produced 
by confinement-related gauge fluctuations rather than instantons.

\bigskip

\end{abstract}
\pacs{}

\vfill

\newpage

\section {Introduction}

In the late 1970's it became clear through the study of instantons
\cite{instantons,tHooftInstantons,Callan1978,ILM} that gauge field 
topology plays an important role in the resolution of the $U_A(1)$ 
problem: topological transitions could produce an $\eta'$ mass via the axial 
anomaly.  It is now quite well established from lattice QCD that topological 
charge is indeed important, with the $\eta'$ mass quite nicely explained by 
its coupling to the $U_A(1)$ current anomaly and with topological charge
playing a significant role in other parts of QCD as well
\cite{Diakonov_et_al,Shuryak_et_al,other_recent_instantons,Schafer_Shuryak,lattice_instantons,DISinstantons,instanton_hadronic_models,instantonsandconfinement,Callan1978merons,Teper}.

In two remarkable 1979 papers \cite{WittenUA(1)}, Witten argued that,
while topological charge fluctuations were clearly involved, the {\it dynamics}
underlying the $\eta'$ mass need not be associated with the semiclassical 
tunnelling events called instantons since the large vacuum fluctuations associated
with confinement would also necessarily entail large fluctuations of topological charge. 
In fact Witten argued that not only are
instantons not required for resolving the $U_A(1)$ problem, but that the instanton
resolution is in conflict with predictions based on the large-$N_c$ approximation.
Instantons would produce an $\eta'$ mass that vanishes exponentially for large $N_c$,
while considerations based on large-$N_c$ chiral dynamics strongly suggest that the $\eta'$
mass should be of order $1/N_c$. Given the incompatibility of large-$N_c$ dynamics and
instantons, and the strong empirical support for the applicability of the
large-$N_c$ limit, Witten speculated that the true dynamical origin of
the $\eta'$ mass would be the coupling of the $U_A(1)$ anomaly to topological
charge associated with confinement-related vacuum fluctuations and {\it not} instantons.

From this point of view, the particular kind of quantized, locally self-dual or anti-self-dual
gauge fluctuations associated with an instanton gas or liquid model of the QCD vacuum are only a 
small subset of all the configurations containing significant fluctuations of topological
charge. Although instantons are favored by the Yang-Mills action, apparently
the much greater entropy 
of the many non-instanton configurations causes them to dominate the QCD path integral. 
As a result, the instanton expansion is invalidated, presumably by large quantum fluctuations associated
with confinement. A very 
analogous situation in the two-dimensional CP(N-1) model is cited by Witten as further 
evidence that even a theory with classical instanton solutions can have a quantum mechanical 
ground state which is not even approximately described by an instanton 
expansion~\cite{Jevicki}. Witten's arguments are also consistent with a view of the QCD vacuum 
based on the strong-coupling expansion in which the gauge fields are highly disordered. 
This disorder leads very directly to confinement, as seen from the area law exhibited by 
large Wilson loops. The semiclassical gauge configurations associated with an instanton 
picture are highly ordered and quite atypical in a strong-coupling framework. 

In this paper we investigate the prevalence of instantons in the QCD vacuum using numerical
lattice QCD. In the absence of analytic control in the continuum, this is the most
direct way of resolving this nonperturbative issue from first principles. To achieve this 
it is necessary to design a test that (a) is capable of distinguishing the typical vacuum
configurations implied by the instanton picture from the bulk of generic configurations
containing topological charge fluctuations, and (b) can lead to a meaningful result using
available lattice techniques. With respect to (a) we propose to test the local
(anti)self-duality properties of the gauge fields in the regions where the field
strength is large. Indeed, it is inherent to an instanton picture of the gauge vacuum
(such as the Instanton Liquid Model (ILM)~\cite{ILM}), that strong fields are concentrated in
small regions of space-time, with the fields being locally (anti)self-dual. Checking
gauge field duality properties directly on Monte-Carlo generated lattice gauge configurations is
problematic because, as is frequently pointed out, the short distance fluctuations associated
with lattice artifacts can mask the expected continuum-like behavior. To avoid artificially
changing the configuration in some type of smoothing or cooling procedure, we use
instead fermionic methods. In particular, we will argue that local
self-duality properties of gauge fields are encoded in the {\it local chirality} properties
of low-lying Dirac eigenmodes. {\it The main point of this paper is to present evidence
that the local chiral structure of low-lying Wilson-Dirac eigenmodes is not consistent
with instanton dominance.} 

The gauge configurations used in this study are a subset of the Fermilab ACPMAPS b-ensemble recently 
used in extensive calculations~\cite{chlogs,scprop} which confirmed in considerable detail the 
chiral properties expected from continuum chiral Lagrangian arguments. These included 
the magnitude and time-dependence of the $\eta'$ hairpin correlator, the value of the 
pure-glue topological susceptibility, and a number of ``quenched chiral logarithm'' 
effects associated with the $\eta'$. Since these previously reported results also probe the chiral 
structure of low-lying Dirac eigenmodes, they provide us with reasonable confidence that 
the qualitative conclusions reported here represent the behavior of continuum QCD. 

Given the conclusion that instantons are not the dominant source of topological charge 
fluctuations, how does this affect our picture of low energy QCD?
Instantons are usually invoked in qualitative explanations
of phenomena associated with topological charge. However, as was explained by 
Witten~\cite{WittenUA(1)}, these physical effects do not necessarily require instantons, 
and can be qualitatively understood in any picture involving confinement. (Note that it has
long been recognized that instantons alone are not likely
to lead to confinement, so it is natural to suspect that their relevance 
to low energy QCD is limited.) An important implication of our result is that it calls into 
question phenomenological approaches based on semiclassical methods in favor of
approachs where confinement plays a central role. In contrast to the situation
for instantons, there is no all-encompassing confinement-based framework comparable to the ILM,
providing a physics picture for how the key low energy phenomena of QCD are driven by
confinement. However, in this paper we will refer to the new
Strong Coupling Quark Model~\cite{SCQM} as possibly providing such a framework. 
In his discussion of these matters, Witten used not only the generic properties of the confining vacuum, but 
also specific properties following from the large-$N_c$ expansion of 
QCD~\cite{LargeNctHooft,LargeNcWitten}. The derivation of the Witten-Veneziano 
formula~\cite{WittenVeneziano} for the $\eta'$ mass serves as an example of how the large-$N_c$
limit, can be used as a tool in quantitative
calculations relevant to these questions.

In fact, the success of the large-$N_c$ expansion is in itself evidence for a non-instanton 
picture of the QCD vacuum.
In this regard, we remind the reader that the successes of large $N_c$ 
are by now very broad and include a basis for the valence quark model~\cite{IsgPat,SCQM}, Regge 
theory~\cite{DTE}, the utility of the narrow resonance approximation, quasi-two-body dominance 
of hadronic decays, the OZI rule~\cite{Zweig,otherOZI,GIonOZI,IsgurThacker}, and the general
systematics of hadronic spectra and matrix elements~\cite{ManoharJenkins,Lebed}.
Also, a growing body of lattice results appear to confirm the empirical evidence 
that the large-$N_c$ expansion is good, {\it i.e.} that $1/N_c=1/3$ is reasonably 
close to $1/N_c=0$. Most directly relevant to the $\eta'$ mass problem are the 
lattice results showing that the topological suceptibilities $\chi_t$ at $N_c=2$ 
and $N_c=3$ are almost equal within errors. With the string tension $\sigma$ used to set 
the scale, it is found \cite{TopSusc} that
${\chi^{1 \over 4}_t / {\sqrt \sigma}}=0.455 \pm 0.015$ 
in $SU(3)_c$ while this same dimensionless ratio takes the value
$0.487 \pm 0.012$ in $SU(2)_c$. Finally, it should be pointed out that Witten has recently
provided new evidence in favor of his conjecture
on the purely theoretical front~\cite{WittenAdS}. Using the 
Anti-de Sitter space-conformal field theory (AdS/CFT) connection which has emerged from 
superstring theory, and assuming that there are no 
phase transitions as a certain parameter $\eta$ varies between the region 
$\eta \rightarrow 0$ (where it approaches 4D gauge theory) and $\eta >>1$ (where it becomes 
a weakly coupled string theory which can be studied as a long wavelength supergravity 
theory), Witten showed that the $\theta$-dependence of $SU(N_c)$ gauge
theory is of leading order in $N_c$ as 
$N_c \rightarrow \infty$, as anticipated~\cite{WittenUA(1)}. In an instanton picture, 
the $\theta$-dependence would vanish exponentially. The AdS/CFT correspondence also reveals
the existence of stable nondegenerate vacua and nonanalytic $\theta$-dependence of the vacuum
energy in the large $N_c$ limit, confirming other aspects of Witten's no-instanton hypothesis.

In Section~\ref{sec:witten} we begin by introducing the reader to the circle of ideas that 
lead Witten to conclude that at the quantum level, instantons might simply ``disappear'' 
from QCD~\cite{WittenUA(1)}. This includes an elementary discussion of how the physical 
effects usually ascribed to instantons naturally occur in a confining vacuum 
even without instantons. In Section~\ref{sec:modes} we discuss the relation between Dirac 
near-zero modes, the chiral condensate, and the $\eta'$ mass. Sections~\ref{sec:witten} 
and~\ref{sec:modes} both have a pedagogical character and serve as theoretical background 
for Section~\ref{sec:locchiral}, where we present the main results of the paper. There we show how an 
instanton-dominated vacuum would be reflected in the local chiral structure of near-zero 
modes, describe our lattice methods, and present the results. Finally, 
Section~\ref{sec:conclude} summarizes our conclusions.

\bigskip 

\section {Witten's Conjecture}
\label{sec:witten}

\subsection{The Vulnerability of QCD Instantons}
\label{subs:wittenevaporate}

To illustrate how large quantum corrections could make instantons irrelevant 
(``evaporate'') in QCD, Witten considered two $U(1)$ gauge theories in $1+1$ 
dimensions, namely
\begin{equation}
{\cal L}_1 \;=\; D_{\mu}\phi^* D_{\mu} \phi - 
                {\mu}^2 \phi^* \phi - {1 \over 4} F_{\mu\nu} F_{\mu\nu}
 \label{eq:40}
\end{equation}
\begin{equation}
{\cal L}_2 \;=\; D_{\mu}\phi^* D_{\mu} \phi - 
                 \lambda( \phi^* \phi-a^2)^2 - 
                 {1 \over 4} F_{\mu \nu} F_{\mu \nu}~,
 \label{eq:80}
\end{equation}
where $D_{\mu}\equiv\partial_{\mu} + i e A_{\mu}$, $\phi$ is a charge scalar
field, $F_{\mu \nu}=\partial_{\mu}A_{\nu}-\partial_{\nu}A_{\nu}$, and the theories are
defined in an infinite volume. The first theory has an unbroken $U(1)$ 
symmetry and a spectrum of neutral $\phi \bar \phi$ bound states (since in 
one dimension the Coulomb potential is a linear confining one). The second 
has a characteristic Higgs spectrum with a real physical scalar and a massive 
``photon". Both ${\cal L}_1$  and ${\cal L}_2$  possess the topological 
charge $Q={e \over {2\pi}}\int d^2x \epsilon_{\mu \nu} F_{\mu \nu}$; in
${\cal L}_2$ it is quantized to integer values and an instanton gas plays an 
important role, while in ${\cal L}_1$ it is not quantized and instantons play 
no role. Of course, even in theory (\ref{eq:40}) there is a global quantization 
of Euclidean topological charge on a finite periodic torus, but in this case the quantization 
is strictly a finite volume effect (analogous to the quantization of momentum 
in a finite box) and has no relevance to the local structure of the vacuum.

Another distinguishing feature of the non-instanton theory 
is in the dependence of physics on the vacuum 
parameter $\theta$ (Fourier conjugate to the topological charge). 
For the spontaneously broken theory (\ref{eq:80}), $\theta$-dependence arises via an instanton 
expansion and physical quantities are smooth analytic functions of $\cos\theta$, 
while for the unbroken theory (\ref{eq:40}), $\theta$ represents a background
electric field, and physical quantities (e.g., the vacuum energy) are singular at
certain values of $\theta$, corresponding to the physical effect of a $\phi^+\phi^-$
pair popping out of the vacuum and screening and cancelling a unit of electric flux.
Witten's recent consideration of this issue in 4D gauge theory via the AdS/CFT
correspondence \cite{WittenAdS} has demonstrated that a similar nonanalytic behavior 
of the vacuum energy as a function of $\theta$ arises in the large $N_c$ limit of QCD
(with wrapped sixbranes playing a role analogous to units of electric flux in the 2D case). 
Thus, at least at large $N_c$, the $\theta$ dependence of four-dimensional $SU(N_c)$ 
gauge theory resembles unbroken 2D scalar electrodynamics~(\ref{eq:40}), and not 
the instanton dominated Higgs theory~(\ref{eq:80}).  

Although it is not believed to happen in two-dimensional scalar electrodynamics, it is 
easily imaginable that a theory with classical instantons, like ${\cal L}_2$, could
be converted into a theory like ${\cal L}_1$ by quantum corrections, if those corrections
changed the broken Higgs-type effective potential to an unbroken one. Given the apparent 
incompatibility of instanton and large-$N_c$ ideas and the strong phenomenological 
evidence for the latter, Witten conjectures that, in spite of the fact that
QCD at the classical level resembles a theory of type ${\cal L}_2$, quantum effects most
likely change it into a theory of type ${\cal L}_1$. To illustrate this view more
convincingly, he then argues in detail that precisely this kind of behaviour occurs
in the two-dimensional CP(N-1) model. This model has a global SU(N) 
symmetry and, like QCD, is asymptotically free and classically scale invariant. 
The condition of finite action leads to a field boundary condition at infinity 
that breaks the SU(N) symmetry, and results in a quantized topological charge 
at the classical level. There are instantons for all N, but the solution 
of the theory in terms of a 1/N expansion reveals that instantons in fact 
evaporate~\cite{Jevicki}. This can be interpreted as invalidation, by large quantum 
corrections, of the boundary condition that breaks the SU(N) symmetry. In the true 
vacuum, the symmetry is actually restored.    

At the classical level, the situation in QCD is quite analogous to
that of a CP(N-1) model. Furthermore, Witten argues, the phenomenon of 
confinement in QCD plays an analogous role to that of an SU(N) symmetry 
restoration in the CP(N-1) case. Indeed, the standard semiclassical condition of 
finite action results in the requirement that the field is pure gauge 
at infinity. However, with the assumption of confinement, this condition 
is not reasonable in QCD. The correct boundary condition at infinity should reflect the
behaviour that is typical of a confining vacuum, while the pure gauge behaviour 
corresponds to a perturbative, non-confining vacuum. 
This suggests that the QCD path integral is dominated by configurations which
invalidate the semiclassically 
motivated pure gauge boundary condition, and that the instantons ``evaporate'' 
due to large quantum corrections.

\bigskip

\subsection{Low Energy QCD without Instantons}
\label{subs:wittenconfinement}

An important role in the qualitative arguments of Subsection II.A
is clearly played by the assumption of confinement. While the structure of the 
confining vacuum is not fully understood, Witten argued that large gauge field 
fluctuations, generically present in the confining vacuum, are sufficient for a qualitative 
explanation of the physical effects usually associated with instantons. This is quite 
easily seen, since many of these effects, including the $\theta$ dependence of physics 
and the solution of the $U_A(1)$ problem, depend only on the fact that, for the configurations 
typical of the QCD vacuum, topological charge fluctuations can be large (or, more precisely, 
that the topological susceptibility of the pure-glue vacuum is nonzero). The Euclidean 
topological charge is defined as
\begin{equation}
   Q\,\equiv\, {1\over{32\pi^2}}\int d^4x \, 
      {1\over 2}\epsilon_{\mu\nu\rho\sigma}G_{\mu\nu}^a G_{\rho\sigma}^a
    \,\equiv\, {1\over{32\pi^2}}\int d^4x \,
      G_{\mu\nu}^a {\tilde G}_{\mu\nu}^a
\end{equation} 
where $G_{\mu\nu}^a = \partial_{\mu} A_{\nu}^a - \partial_{\nu} A_{\mu}^a
       + f^{abc} A_{\mu}^b A_{\nu}^c$. A basic property 
of $Q$ is that its density is a total divergence and thus it can be represented 
as an integral over the large hypersphere
\begin{equation}
  Q \,=\, \int d^4x\, \partial_{\mu}K_{\mu}  
    \,=\, \int dS_{\mu}\, K_{\mu}
    \label{eq:200}
\end{equation}
with $dS_{\mu}$ the surface element, and 
\begin{equation}
   K_{\mu} \,=\, {1\over{16\pi^2}}\, \epsilon_{\mu\nu\rho\sigma}\,
                 \biggl(\,A_{\nu}^a\partial_{\rho}A_{\sigma}^a +
                 {1\over 3} f^{abc} A_{\nu}^aA_{\rho}^bA_{\sigma}^c\, \biggr)
     \label{eq:240}
\end{equation} 
the topological current.

    Let us now consider the $\theta$ dependence of physics. Before the 
discovery of instantons, the inclusion of the perfectly acceptable term 
$S'\propto \theta\int d^4x\, G_{\mu\nu}^a {\tilde G}_{\mu\nu}^a$ into the gauge 
action was believed to have no effect on physical observables, since,
according to (\ref{eq:200}) it is just a surface term. The behaviour of fields 
at infinity should reflect the ``typical behaviour in the vacuum'' which, 
according to usual (perturbative) lore inherited from QED, corresponds to 
physical fields being zero (pure gauge). Hence, it appeared that there was no 
reason to believe that the $\theta$ term would contribute, and $S'$ 
was discarded. 

    The discovery of instantons revealed that even if one retains the picture of 
the vacuum as asymptotically pure gauge, the integral 
$\int d^4x \,G_{\mu\nu}^a {\tilde G}_{\mu\nu}^a$ does not 
necessarily have to vanish. Indeed, the BPST instanton
solution~\cite{instantons}, even though pure gauge at infinity, winds nontrivially around the group
and yields a nonzero value for the integral ($Q=1$). More generally, any gauge
configuration that is pure gauge at infinity (not just solutions of the 
equation of motion) and winds nontrivially around the group will have a
nonzero, integer value of Q.
Since the number of windings is stable under a continuous deformation 
of gauge fields, this pointed to the role of topology in QCD and, more importantly
lead to the fundamental discovery that physics actually will depend on $\theta$
(except in the presence of massless fermions).      

At the same time, from the confinement point of view, $\theta$ dependence
is rather obvious. If confinement is assumed, one simply should not describe 
the vacuum as being asymptotically pure gauge. Consequently, the correct boundary 
condition at infinity should involve a nonzero $G_{\mu\nu}^a$ in some form. 
For such configurations there is absolutely no reason why 
$\int d^4x \,G_{\mu\nu}^a {\tilde G}_{\mu\nu}^a$ should vanish, hence 
the $\theta$ dependence. A concrete model which illustrates that instantons 
do not have to play any significant role in this respect is the Schwinger model, 
which is a theory of type (\ref{eq:40}) without quantized topological charge. 
The $\theta$ term corresponds to a background electric field and the physics 
depends on it, except when fermions are massless.    

By the same argument, confinement also provides an alternative mechanism for $\eta'$
mass generation.
The flavor singlet axial current 
$j^{\mu}_5 \equiv \Sigma_{\alpha} 
           {\bar\psi}_{\alpha} \gamma^{\mu} \gamma_5 \psi_{\alpha}$ ($\alpha$
being the flavour index) exhibits an axial anomaly, but the corresponding 
divergence is proportional to the topological charge density, 
$\partial_{\mu}j^{\mu}_5 \propto G_{\mu\nu}^a {\tilde G}_{\mu\nu}^a$. 
Consequently, one can use the topological current (\ref{eq:240}) to define 
a conserved, but gauge variant current. To solve the $U_A(1)$ problem, it is
necessary to show that this conserved current does not lead to a Goldstone
pole~\cite{KogutSusskind}, so that the $\eta'$ remains massive in the chiral limit, unlike 
the other pseudoscalar mesons. t'Hooft has shown~\cite{tHooftInstantons} that in 
the semiclassical approximation, using the instanton gas picture, this is indeed 
what happens. At the same time, as Witten argues, this conclusion only relies on 
the fact that the semiclassical framework naturally emphasizes configurations that 
give nonzero $\int d^4x \,G_{\mu\nu}^a {\tilde G}_{\mu\nu}^a$. While it is 
interesting that the $U_A(1)$ problem is solved already at the semiclassical level, 
the fluctuations of a confining vacuum, as we noted above, will also naturally 
include configurations with nonvanishing topological charge. As a result,
the $U_A(1)$ problem is expected to be solved in any confinement-based 
picture of the QCD vacuum. This is exhibited, for example, by the derivation 
of the Witten-Veneziano formula~\cite{WittenVeneziano} for the $\eta'$ mass, 
which is carried out in the large $N_c$ framework without reference 
to semiclassical arguments.

\subsection{Discussion}
\label{subs:wittendiscussion}

Early attempts at studying topological effects in QCD were framed in
the context of a dilute instanton gas.
While it is now well known that this framework is not capable 
of providing an accurate description of low energy QCD, the associated basic 
picture of the vacuum has survived in more elaborate scenarios based on
instanton methods, such as the Instanton Liquid Model~\cite{ILM}. In particular, 
the typical vacuum gauge configuration is imagined as being very inhomogeneous, 
containing lumps of very strong (anti)self-dual fields, surrounded by regions of 
very weak fields. When the non-interacting instanton gas picture is refined to include 
correlations, one can no longer speak of first-principles semiclassical calculations, 
but rather of a semiclassically-motivated phenomenology.  
An extensive examination of the possible roles of instantons in low-energy QCD 
(including their role in spontaneous chiral symmetry breaking and the propagation
of light quarks in the vacuum) has 
been made in the context of the ILM~\cite{ILM}. Ref.~\cite{ILM} also provides an excellent 
overview of the  history of instantons and some very beautiful pedagogical discussion of 
low energy QCD.
For other recent important work on instantons, see
\cite{Diakonov_et_al,Shuryak_et_al,other_recent_instantons,Schafer_Shuryak,lattice_instantons,DISinstantons,instanton_hadronic_models}. 

To further motivate our conclusion that such models are not viable, we note 
that from the confinement point of view, 
the instanton picture is unnaturally ordered. Indeed, with the (anti)self-dual restriction 
entailed by instanton excitations, the chromoelectric and chromomagnetic fields are locked 
together $({\bf E^a}=\pm{\bf B^a})$. The important property of this condition is that the role 
of topological charge is maximally emphasized in the sense that for a given value 
of the action density, (anti)self-duality implies the maximal topological 
charge density. This can be seen from the fact that
$S\propto ({\bf E}\cdot{\bf E} + {\bf B}\cdot{\bf B})$, while
$Q\propto {\bf E}\cdot{\bf B}$. The resulting lumpiness of topological charge
then affects the propagation of quarks. While the large value of
topological susceptibility in the pure gauge vacuum obtained from lattice
simulations suggests that topological charge is indeed lumpy~\cite{Teper},
this lumpiness is perfectly natural for a highly fluctuating confining
vacuum as well: confinement automatically entails large localized
fluctuations of ${\bf E^a}$ and ${\bf B^a}$ with energy densities 
of the order of $1 GeV/fm^3$, generating ``hot spots'' in both
$({\bf E}\cdot{\bf E} + {\bf B}\cdot{\bf B})$ and ${\bf E}\cdot{\bf B}$.
What is not natural from the confinement point of view is the maximally
ordered situation implied by the instanton picture. There appears to be no
good reason to expect that the confining vacuum would fluctuate in such
a way as to only generate (anti)self-dual lumps. In the SCQM~\cite{SCQM},
these fluctuations are associated with virtual scalar glueballs mixed into
the bare strong-coupling vacuum (leading to a strong space-time localization
of the gauge field fluctuations) and a loosely correlated angle between
${\bf E^a}$ and ${\bf B^a}$ associated with the fields internal to the
scalar glueball.
 
\section {The Role of Fermionic Near-Zero Modes}
\label{sec:modes}
 
Since the approach pursued in this paper relies on fermionic near-zero
modes, we now discuss their role in determining the low energy properties
of QCD. In particular, we will concentrate on spontaneous chiral symmetry 
breaking (s$\chi$SB) and the $\eta'$ mass. The common origin of these phenomena 
is an attractive feature of the ILM, but we will argue below that the same 
connection will arise in any picture which incorporates large fluctuations of 
topological charge density. 

We begin by recalling the derivation of the Banks-Casher relation~\cite{BanksCasher}
which clarifies the role of near-zero modes of the Dirac operator in forming the 
chiral condensate. (Although we will ultimately address these issues numerically,
using the lattice Wilson-Dirac operator, we restrict our theoretical considerations to 
the continuum Dirac operator. The effects of lattice discretization will be discussed
in the next Section in the context of the Monte Carlo results.) 
The spectral requirement for a nonvanishing chiral condensate is found by considering 
the eigenmode expansion of a scalar quark loop with mass $m$ in a background gauge 
field $A$ in Euclidean 4-space, 
\begin{equation}
\label{eq:scalarloop}
\langle \bar{\psi}(x)\psi(x)\rangle_A \,=\, {\rm Tr}\,G(x,x) \,=\, 
\sum_i\frac{\psi_i^{\dag}(x)\psi_i(x)}{-i\lambda_i-m}\;\;.
\end{equation}
Here the $\psi_i$'s and $\lambda_i$'s are the eigenvectors and eigenvalues of the
{\it massless} Dirac operator,
\begin{equation}
{\mbox{$i\,\!\!\not\!\!D$}} \psi_i \,=\, \lambda_i\psi_i
\end{equation}
Because of the anticommutator  
$\{\gamma^5,\,\,{\mbox{$\!\!\not\!\!D$}} \}=0$, 
nonzero eigenvalues come in chiral pairs $\pm\lambda_i$ with eigenvectors 
$\psi_i$ and $\gamma^5\psi_i$, respectively. Symmetrizing the eigenvalue 
expansion (\ref{eq:scalarloop}) over chiral pairs, we have
\begin{equation}
\langle \bar{\psi}(x)\psi(x)\rangle_A \,=\,  
-\sum_{\lambda_i>0}\left(\frac{2m}{\lambda_i^2+m^2}\right)\psi_i^{\dag}(x)\psi_i(x)
- \frac{1}{m}\sum_{\lambda_i=0} \psi_i^{\dag}(x)\psi_i(x)\;\;.
\end{equation}
After averaging over gauge fields, the expectation value on the left hand side
becomes translation invariant, so we may integrate over the 4-volume $V$ and use the
fact that the eigenfunctions are normalized to $\int\psi_i^{\dag}\psi_i d^4x = 1$ 
to get
\begin{equation}
\langle \bar{\psi}\psi\rangle = -\frac{1}{V}\sum_i\frac{m}{\lambda_i^2+m^2}
-\frac{N_0}{mV} \;\;.
\end{equation}
At this point it is important to take the volume to infinity {\it before} taking the
chiral limit $m\rightarrow 0$. If we were to take the chiral limit first, we would
conclude that only the modes with {\it exactly} zero eigenvalue contribute to 
$\langle \bar{\psi}\psi\rangle$. Instead, the infinite volume limit at small but 
finite mass replaces the sum by an integral over a density function $\rho(\lambda)$,
which receives contributions from a large number of near-zero modes. In fact, the
contribution of exact zero modes to the integral becomes negligible compared to the
near-zero modes in the infinite
volume limit. (For constant topological susceptibility,
the typical number of exact zero modes $N_0$ grows like $\sqrt{V}$, while the total
number of low-lying modes grows linearly with the volume.)
As a result, only the density of near-zero modes is relevant for chiral symmetry breaking,
\begin{equation}
\langle \bar{\psi}\psi\rangle = 
      -\int_0^{\infty} d\lambda\,\rho(\lambda)\left(\frac{m}{\lambda^2+m^2}\right)
\end{equation}
In the chiral limit, this reduces to the Banks-Casher relation
\begin{equation}
\langle \bar{\psi}\psi\rangle = -\pi\rho(0)
\end{equation}
where $\rho(0)= \lim_{\lambda\rightarrow 0} \rho(\lambda)$.
Thus, the requirement for spontaneous chiral symmetry breaking is that the density of 
near-zero modes $\rho(0)$ in a very large box must be finite. The number of low-lying 
modes needed to achieve this is of course much larger than what would be available for free fermions 
in four dimensions, where the density vanishes like $\lambda^3$. In the instanton liquid 
model, the required excess of near-zero eigenmodes is posited to be supplied by the topological zero-modes 
of the instantons and anti-instantons, which mix and spread into a band of eigenvalues 
near zero. But a finite value of $\rho(0)$, and hence a chiral condensate, can arise under 
much more general circumstances. The instanton liquid model populates the vacuum with a 
finite density of gauge field ``lumps'' which act like attractive potentials in the quark 
eigenvalue problem. In this case, each lump supports one low-lying eigenmode, so the 
total number of low-lying modes grows linearly with the volume and the finite density
of states required for $s\chi SB$ can be achieved. But note that the Atiyah-Singer index 
theorem~\cite{AtiyahSinger} tells us that any gauge configuration with an overall topological 
charge (which is an integer in a periodic box, with or without instantons) will exhibit at 
least that number of zero eigenvalues in its Dirac spectrum. More generally, large gauge 
fluctuations of all types will act as attractive potentials in the Dirac operator 
(see Section~\ref{sec:locchiral}), giving rise to low-lying ``bound state'' eigenmodes. 
Just as in the instanton model, for a given density of large gauge fluctuations we would 
expect the number of such modes to grow linearly with the volume. This argument 
shows that a finite density of near-zero Dirac eigenmodes, and hence $s\chi SB$, can 
be a result of the types of large gauge fluctuations which are implied by confinement and 
do not require any of the more specific restrictions of the instanton picture.

Given the finite density of low-lying Dirac eigenvalues implied by $s\chi SB$ and the
Banks-Casher relation, we now want to consider the resolution of the $U_A(1)$
problem from the point of view of low Dirac eigenmodes. The most direct way to
address this issue is to consider the eigenmode expansion for the pseudoscalar
``double-hairpin'', {\it i.e.} the loop-loop correlator associated with the $\eta'$ mass,
\begin{equation}
\Delta^{(A)}_{dh}(x,y) = \langle\, {\rm Tr}\,\gamma^5 G(x,x) \, 
                                   {\rm Tr}\,\gamma^5 G(y,y) \,\rangle_A
\end{equation}
As before, the eigenmode expansion represents the quark propagators in a particular 
background gauge configuration $A$. In the end, the expressions are averaged over an 
ensemble of gauge configurations. As in the derivation of the Banks-Casher relation, 
each loop can be symmetrized over the contribution of chiral pairs, giving
\begin{equation}
\label{eq:hp}
\Delta^{(A)}_{dh}(x,y) = \sum_{i,j}\left(\frac{m}{\lambda_i^2+m^2}\right)
\left(\frac{m}{\lambda_j^2+m^2}\right)\psi_i^{\dag}(x)\gamma^5\psi_i(x)
\;\psi_j^{\dag}(y)\gamma^5\psi_j(y)
\end{equation}
Because of the two factors in parentheses, we see that, just like 
$\langle\bar{\psi}\psi\rangle$,
the pseudoscalar hairpin correlator is dominated by the low eigenmodes in the chiral 
limit $m\rightarrow 0$. The size and space-time dependence of this correlator, after
averaging over gauge configurations, can be estimated by chiral Lagrangian arguments.
In either quenched or full QCD, the hairpin correlator will be large and will have a long range
component in the chiral limit which falls off exponentially according to the pion mass.
In quenched QCD, the hairpin correlator at zero 3-momentum falls off 
like $te^{-m_{\pi}t}$, while in full QCD it falls off like $e^{-m_{\pi}t}$ (cancelling 
the Goldstone pole of the valence propagator). 

A recent study of the hairpin correlator~\cite{chlogs}, using the same (quenched) gauge 
configurations that we study here, has confirmed that both the size and time-dependence 
of the hairpin correlator are in excellent agreement with chiral Lagrangian predictions, 
in which the double hairpin vertex is treated as an $\eta'$ mass insertion. The study 
in Ref.~\cite{chlogs} uses a different method for computing the hairpin 
correlator ({\it i.e.} not an eigenmode expansion), but we take those results as strong evidence 
that the correct physical mechanism for $\eta'$ mass generation, in particular the 
structure of the low eigenmodes which should dominate the result, is well represented 
by the lattice data. The interpretation of these results is discussed in both
Ref.~\cite{IsgurThacker} and \cite{SCQM}, where the $\eta'$ double hairpin is an ordinary
$q\bar{q}\rightarrow q'\bar{q}'$ OZI-violating diagram of the quark model.

Before discussing the lattice results of this paper, some further remarks on the eigenmode formula 
for the hairpin correlator (\ref{eq:hp}) will be useful. 
A finite density of modes near $\lambda\approx 0$ is already guaranteed by the existence 
of the chiral condensate. The additional property required to have a large hairpin 
correlator (and hence a finite $\eta'$ mass) in the chiral limit is that the typical 
low-lying eigenmodes must have significant space-time regions in which the pseudoscalar 
charge $\psi_i^{\dag}\gamma^5\psi_i$ is large. Note that, in the Banks-Casher relation, 
the contribution of a given mode to the chiral condensate was fixed by the wave function 
normalization condition $\int d^4x\, \psi_i^{\dag}\psi_i = 1$. By contrast, the pseudoscalar 
charge for a mode $\psi_i$ with nonzero eigenvalue must integrate to zero,
\begin{equation}
\label{eq:orthog}
\int d^4x\, \psi_i^{\dag}\gamma^5\psi_i = 0
\end{equation}
since $\psi_i$ and $\gamma^5\psi_i$ have opposite eigenvalues of
${\mbox{$i\,\!\!\not\!\!D$}}$ and are therefore orthogonal
to each other. What is required to produce a non-vanishing $\eta'$ mass and resolve 
the $U_A(1)$ problem is for the low-lying eigenmodes to exhibit local regions
in which $\psi_i^{\dag}\gamma^5\psi_i$ is large, even though the integrated quantity
(\ref{eq:orthog}) vanishes. This does not happen, for example, for the eigenmodes 
of a free fermion in a periodic box, for which the pseudoscalar product vanishes 
locally, $\psi_i^{\dag}\gamma^5\psi_i = 0$. Eigenmodes which resemble free plane wave 
states will thus not contribute to the $\eta'$ hairpin correlator. In the instanton 
model, the low eigenmode wave functions will exhibit local peaks on the instantons and 
anti-instantons in which $\psi_i^{\dag}\gamma^5\psi_i = \pm \psi_i^{\dag}\psi_i$, 
respectively. Thus, if the instanton picture is correct, the low eigenmodes will produce 
a large hairpin correlator via lumps of right-handed chiral charge 
$\psi_i^{\dag}(1+\gamma^5)\psi_i=\psi_{iR}^{\dag}\psi_{iR}$ 
interspersed with other lumps of left-handed chiral charge
$\psi_i^{\dag}(1-\gamma^5)\psi_i=\psi_{iL}^{\dag}\psi_{iL}$. 
This is a definitive prediction of the instanton model which follows directly from the 
self-duality and anti-self-duality of the gauge lumps, and it is this prediction that 
we address here. What we want to emphasize is that
the resolution of the $U_A(1)$ problem only requires that the low eigenmodes exhibit 
patches or lumps which contain significant amounts of $\psi^{\dag}\gamma^5\psi$ 
charge, not that these lumps be {\it purely} right-handed or left-handed. The view espoused 
by Witten leads to this less restrictive prediction, which is strongly supported 
by our lattice results.

\bigskip

\section {The Chiral Structure of the Near-Zero-Modes}
\label{sec:locchiral}

\subsection {Methods}
\label{subs:lchmethods}

As we discussed in the preceeding section, the structure of the QCD gauge vacuum 
in any instanton-based model involves a particular assertion about the 
structure of typical gauge excitations in the regions in which the field strength is
large. Specifically, these excitations are supposed to be dominantly 
self-dual or anti-self-dual. We will test this assertion by studying the
chiral structure of the low Dirac eigenmodes. In an instanton picture, these
low eigenmodes are approximately given by linear combinations of topological
zero-mode wave functions centered around each instanton and anti-instanton
in the configuration. The fermion zero mode associated with an instanton
(anti-instanton) is a lump of purely right-(left-)handed chiral charge,
$\psi^{\dag}(1\pm\gamma^5)\psi$. For more general large gauge field fluctuations,
we expect that the right- and left-handed components of the low-lying 
eigenfunctions will clump around the same large gauge fluctuations, with
relative strength determined by the relative size of self-dual and 
anti-self-dual gauge components. To see why this is true, consider the eigenfunctions 
of the continuum Dirac operator in a background gauge field. We note that the Dirac 
eigenfunctions satisfy 
${\mbox{$i\,\!\!\not\!\!D$}} \psi \,=\, \lambda\psi$,
from which we can obtain second order differential equations for the left and right 
components,
\begin{equation}
  (\,{\mbox{$i\,\!\!\not\!\!D$}}\,)^2 \psi \,=\, 
  \left[-D^2 + \frac{1}{2}\sigma_{\mu\nu}G_{\mu\nu}\right]\psi = \lambda^2\psi\;\;.
\end{equation}
Separating the gauge field into self-dual and anti-self dual components,
\begin{equation}
G_{\mu\nu} = \frac{1}{2}\left(G_{\mu\nu}+\tilde{G}_{\mu\nu}\right)
+\frac{1}{2}\left(G_{\mu\nu}-\tilde{G}_{\mu\nu}\right)
\equiv G^{(+)}_{\mu\nu}+G^{(-)}_{\mu\nu}\;\;,
\end{equation}
we obtain
\begin{eqnarray}
\label{eq:schrodleft}
\left[-D^2 + \frac{1}{2}\sigma_{\mu\nu}G^{(+)}_{\mu\nu}\right]\psi_L & = &\lambda^2\psi_L \\
\label{eq:schrodright}
\left[-D^2 + \frac{1}{2}\sigma_{\mu\nu}G^{(-)}_{\mu\nu}\right]\psi_R & = &\lambda^2\psi_R\;\;.
\end{eqnarray}
These are 4-D Schrodinger-like eigenvalue equations, with the self-dual and anti-self-dual
components of the gauge field playing the role of a potential term for the left- and
right-handed components of the eigenvector. If the gauge lump is purely self-dual or
anti-self-dual, then only one of the two chiral components will be attracted by the
gauge fluctuation, and the fermion lump will be purely left-handed or right-handed.
For more general gauge fluctuations, we expect the chiral structure of the fermion
lump to vary arbitrarily, as determined by the relative size of the $G^{(+)}$ and 
$G^{(-)}$ components of the gauge lump. 

In the lattice results discussed here, we study the chirality of low Dirac 
eigenmodes by calculating the left and right chiral charges $\psi_L^{\dag}\psi_L$ and 
$\psi_R^{\dag}\psi_R$ for particular eigenvectors in the regions where their wave functions 
are large. Specifically, we scan through each eigenvector site by site and pick out 
the sites for which the overall size of the wave function 
$\psi^{\dag}\psi=\psi^{\dag}_L\psi_L+\psi^{\dag}_R\psi_R$ 
is greater than some minimum. We have chosen this threshold so that we sample 
about $1\% $ of the sites on the lattice for a typical eigenmode. Thus we are looking at 
the very highest peaks of the fermion eigenfunctions. If an instanton-dominated picture 
of low eigenmodes is at all valid, we would expect the peaks of the wave function to 
closely resemble instanton zero modes. Thus, if instantons dominate,
a local peak in the wave function for a 
low-lying eigenmode (a fermion lump) should be dominantly a lump in 
$\psi_L^{\dag}\psi_L$ or $\psi_R^{\dag}\psi_R$, but not both. 
On the other hand, fermion lumps without a definite chirality would be an indication of 
non-self-dual gauge fluctuations, as expected from a confinement-related mechanism.

In the discussion of our results, an important distinction is made between exactly real 
modes and near-real modes of the Wilson-Dirac operator, which correspond to zero modes
and near-zero modes, respectively, of the continuum Dirac operator. In the continuum
theory, an exact zero mode is associated with the global topology of the gauge
field on a 4-torus. In any gauge configuration there would be a minimum number of exact 
zero modes equal to the integrated topological charge.
Such zero modes are unpaired and should be chiral, independent of whether the topological 
charge comes in the form of instantons or some other gauge fluctuations. The true test of 
instanton- vs. confinement-related fluctuations comes from studying the near-real
modes, which also should be {\it locally chiral} if instantons are dominant.
For example, a configuration of a nearby
instanton-anti-instanton pair should produce a pair of near-real eigenmodes, each of which 
exhibit two lumps of opposite chirality. For more general gauge fluctuations which are
not instantons, the attached fermion lumps will not have a definite chirality but will 
have both $\psi_L^{\dag}\psi_L$ and $\psi_R^{\dag}\psi_R$ charges.

\subsection {Results}
\label{subs:lchresults}

Using the procedure outlined in the previous section, we have carried out a detailed 
study of the low-lying eigenmodes of the Wilson-Dirac operator on an ensemble of 
Monte Carlo generated lattice gauge configurations. The gauge configurations used 
were a 30-configuration subset of the b-lattice quenched ensemble from
the Fermilab ACPMAPS library. Low eigenmodes were obtained using the Arnoldi 
algorithm~\cite{ARPACK}. The lattice size is $12^3\times 24$, and the gauge
coupling is $\beta=5.7$ (corresponding to a
lattice spacing of $a^{-1}\simeq 1.18$ GeV).
Although the lattice spacing is relatively coarse, recall that
an extensive study of chiral symmetry and quenched chiral logarithms has been
carried out on this gauge ensemble~\cite{chlogs,scprop},  
showing that this lattice spacing is fine enough to reproduce in remarkable
detail a variety of quenched chiral log and chiral loop effects related to the
$\eta'$ hairpin diagram. Combined with the clear observation of chirality for exactly 
real modes (see below), we are reasonably confident that lattice spacing effects
do not invalidate our conclusions.

As described above, we select a subset of lattice points for each eigenmode, picking 
out the regions in which the wave function is large. For each of the chosen points, we 
calculate the magnitudes of the left and right components and parametrize their ratio by a
``chiral orientation'' parameter $X$, defined by
\begin{equation}
\tan\,\left(\frac{\pi}{4}(1+X) \right) = \frac{|\psi_L|}{|\psi_R|} \;.
\end{equation}
Note that $X$ is just an angle in the $|\psi_L|$-$|\psi_R|$ plane, rescaled to
run between -1 and +1. In particular, if the fermion eigenmode lumps have a
random chiral orientation,
this will be reflected as a flat distribution in $X$. On the other, hand if the
eigenmodes are locally chiral, as with instanton zero-modes, the distribution will
be peaked near $X\approx \pm 1$.
For any particular eigenmode or set of eigenmodes we wish to examine, we construct
a histogram of values of the chirality parameter $X$ for all the space-time points at which
the size of the wave function is above threshold. 
In our examination of 1592 low-lying eigenmodes on 30 gauge
configurations, we found three distinct types of chirality 
histograms: \\
(1) The exactly real (unpaired) Wilson-Dirac eigenmodes in the continuum band ({\it i.e.} near 
$\kappa_c$) show a significant degree of chirality, as seen in Fig.~1,
{\it i.e.} the histogram exhibits clear peaks at the two ends of the histogram 
$X\approx \pm 0.5$. (Lattice effects have shifted these peaks from their continuum location 
of $\pm1$, but the peaks are still clearly visible.) Fig. 1 includes results from 37
real modes.\\
(2) The near-real (paired) modes show a distribution which is {\it strikingly flat} over the
central region of $X$ between $\approx -.5$ and $.5$, as seen in Fig.~2.
This plot shows the chirality distribution for 350 complex modes with imaginary parts 
between 0.004 and 0.1 in lattice units (about 5 MeV to about 120 MeV in physical units). \\
(3) The eigenmodes with larger imaginary parts givea a chirality distribution that is peaked in
the central region $X\approx 0$. 
This is seen in Fig.~3. This plot includes 336 modes with complex eigenvalues that have
$|{\rm Im}\lambda|>0.5$.

We now discuss the significance of each of these three types of chirality distribution.
As we have discussed, the chirality of the exactly real modes is expected from general 
considerations, and has nothing to do with instantons {\it per se}. For continuum Dirac
eigenmodes, $\gamma^5$ reflection relates conjugate pairs of eigenmodes at $\pm\lambda$. 
Thus, an unpaired, nondegenerate mode with zero eigenvalue must be an eigenstate of $\gamma^5$, 
{\it i.e.} $\gamma^5\psi_0(x)=\pm\psi_0(x)$. Of course, this argument is rendered inexact by 
lattice effects and the non-antihermiticity of the Wilson-Dirac operator. Nevertheless, 
the lattice results show clear evidence of the required chirality of the exactly real modes. This 
becomes even more clear if we look at histograms for individual real modes. In nearly all 
cases, the histogram for a single real eigenmode only exhibits one chiral peak, not two, 
as seen in the plots of three typical real modes (taken from three different gauge 
configurations) in Fig.~4. Thus the zero-modes exhibit a {\it global} chirality, as
expected. Again we emphasize that the expectation that a zero-mode will exhibit an overall 
chirality has nothing to do with instantons but is required by general principles: an exact
zero mode will appear in the spectrum if the overall integrated topological charge differs from 
zero~\cite{AtiyahSinger}. In an instanton picture, a positive chirality zero mode will have 
wave function peaks at the locations of some or all of the instantons but will be small 
in the vicinity of the anti-instantons. But chiral zero modes will occur even if
the topological charge appears in the form of non-self-dual ({\it e.g.}, random) gauge fluctuations. 

Next we turn to the near-real modes which come in conjugate pairs and are therefore
not required to be chiral. In fact, their global chirality must vanish as a 
consequence of $\gamma^5$ hermiticity of the Wilson-Dirac operator. The chirality histogram 
for near-real modes is shown in Fig.~2. In assessing our lattice results, we must be aware 
of the possibility that finite lattice spacing effects may obscure the chiral behavior 
expected from continuum arguments. It is therefore particularly reassuring that
we are able to clearly observe chirality peaks in the histogram of exactly real
modes, Fig. 1, where chirality is required on general principles. It is difficult to imagine
lattice effects which would destroy the chirality of the near-real modes without
also destroying it in the exactly real modes. The contrast between the chiral structure 
of the exactly real eigenmodes (Fig. 1 and Fig. 4) and the flat random chirality seen in the near-real modes
(Fig. 2 and Fig. 5), thus makes the latter result quite compelling. One very identifiable lattice effect is that 
the histograms are suppressed near the boundaries $X=\pm 1$, so that, for example the 
chirality peaks for the real modes are at $X\approx\pm 0.5$ instead of $\pm 1$. 
The same effect is visible at the ends of the histogram for the near-real modes, Fig.~2, 
where the flat central plateau falls off rapidly for $|X|>0.5$. Like the exactly real modes, 
the chiral structure of the near-real modes is further clarified by studying histograms 
for individual modes. The histograms for three typical modes are shown in Fig.~5. 
The individual modes displayed in Fig.~5 were intentionally chosen to be particularly 
close to the real axis, with imaginary parts of 5 MeV, 10 MeV, and 16 MeV, respectively. 
The Arnoldi algorithm has no trouble resolving such nearly real eigenvalues into complex
conjugate pairs, so in practice we found no difficulty in distinguishing between
unpaired, exactly real modes, and nearly real, paired modes. The point emphasized 
by the modes shown in Fig. 5, and by the other near-real modes that we have
inspected individually, is that {\it even modes very close to the real axis show no
tendency to be locally chiral}, in marked contrast to the exactly real, unpaired modes.

The flat chirality distribution for the near-real modes exhibited in Fig.~2 is 
our central result. This may be contrasted not only with the chiral peaking of
the real-mode histogram, but also with the central peaking of the higher momentum
modes in Fig. 3 (with eigenvalues $0.5<{\rm Im}\;\lambda<1.0$). 
The central peaking at $X\approx 0$ can be interpreted as approximate ``plane
wave'' behavior, in that these modes exhibit relatively little pseudoscalar charge. 
Note that the flat chiral distribution of the low-lying modes in Fig. 2
corresponds to generally nonvanishing $\psi^{\dag}\gamma^5\psi$ charge in the wave function
peaks. Thus, these modes provide the mechanism for producing the $\eta'$ mass.
To the extent that the higher plane-wave-type modes are centrally peaked,
they have $\psi^{\dag}\gamma^5\psi\approx 0$, and do not contribute significantly to the $\eta'$
mass. Of course, even the modes in Fig. 3 have modest-sized eigenvalues of $\approx 600$ MeV
to a GeV, and so the peak around $X=0$ is still fairly broad. By examining histograms for various 
ranges of eigenvalues, we see a gradual transition from the plateau-like behavior for
small values of Im $\lambda$ to the more centrally peaked distribution of Fig. 3. By
contrast, the transition from Fig. 1 to Fig. 2 is sudden, with no indication of 
chiral peaks even for complex modes very close to the real axis.

\newpage

\vspace{1.0in} 
\begin{figure}\vspace*{2.0cm}
\includegraphics{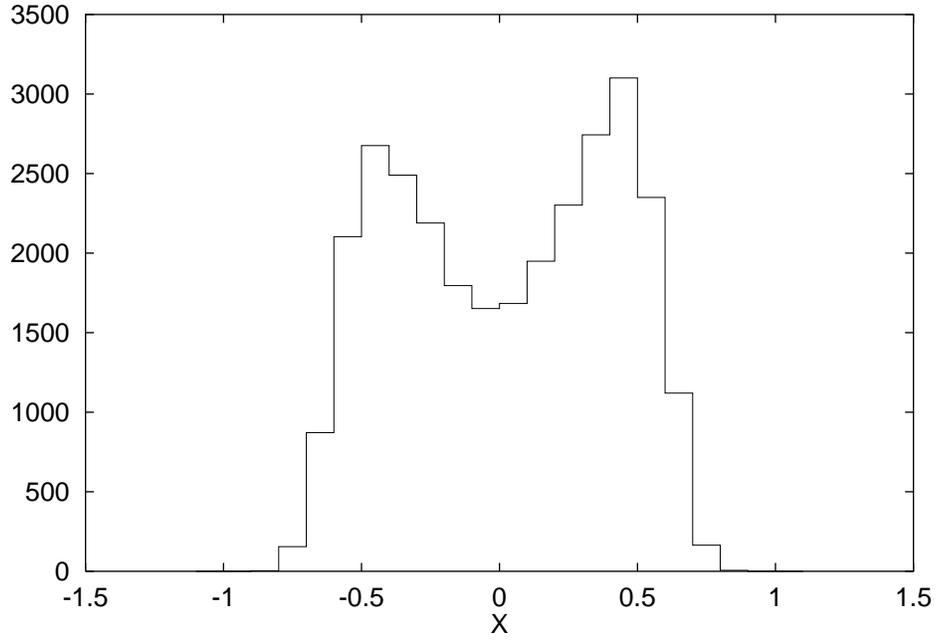}
\vspace{7.0cm}
\caption[]{Chirality histogram for exactly real Wilson-Dirac eigenmodes near $\kappa_c$.  }
\label{fig:real_modes}
\end{figure}

\vspace{0.3in} 
\begin{figure}\vspace*{1.0cm}
\includegraphics{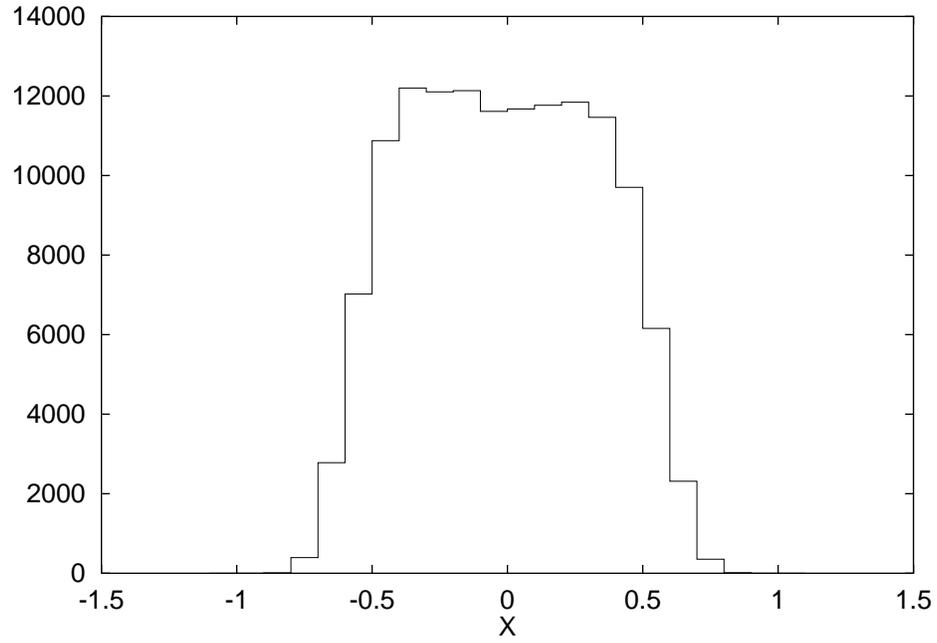}
\vspace{7.0cm}
\caption[]{Chirality histogram for near-real modes with $|{\rm Im} \lambda| < 0.1a^{-1}
\approx 120$ MeV.
  }
\label{fig:near_real_modes}
\end{figure}

\newpage
\vspace{1.0in}
\begin{figure}\vspace*{2.0cm}
\includegraphics{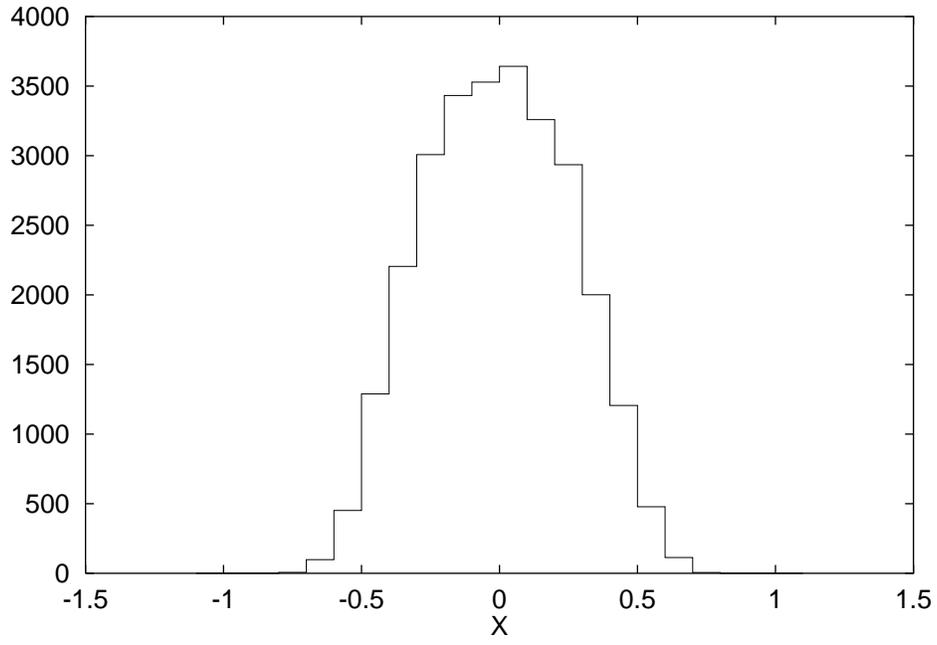}
\vspace{7.0cm}
\caption[]{Chirality histogram for complex modes with $|{\rm Im} \lambda| >0.5a^{-1}\approx
600$ MeV. }
\label{fig:complex_modes}
\end{figure}
\vspace{0.3in}

\newpage

\vspace*{1.0cm}
\begin{figure}
\includegraphics{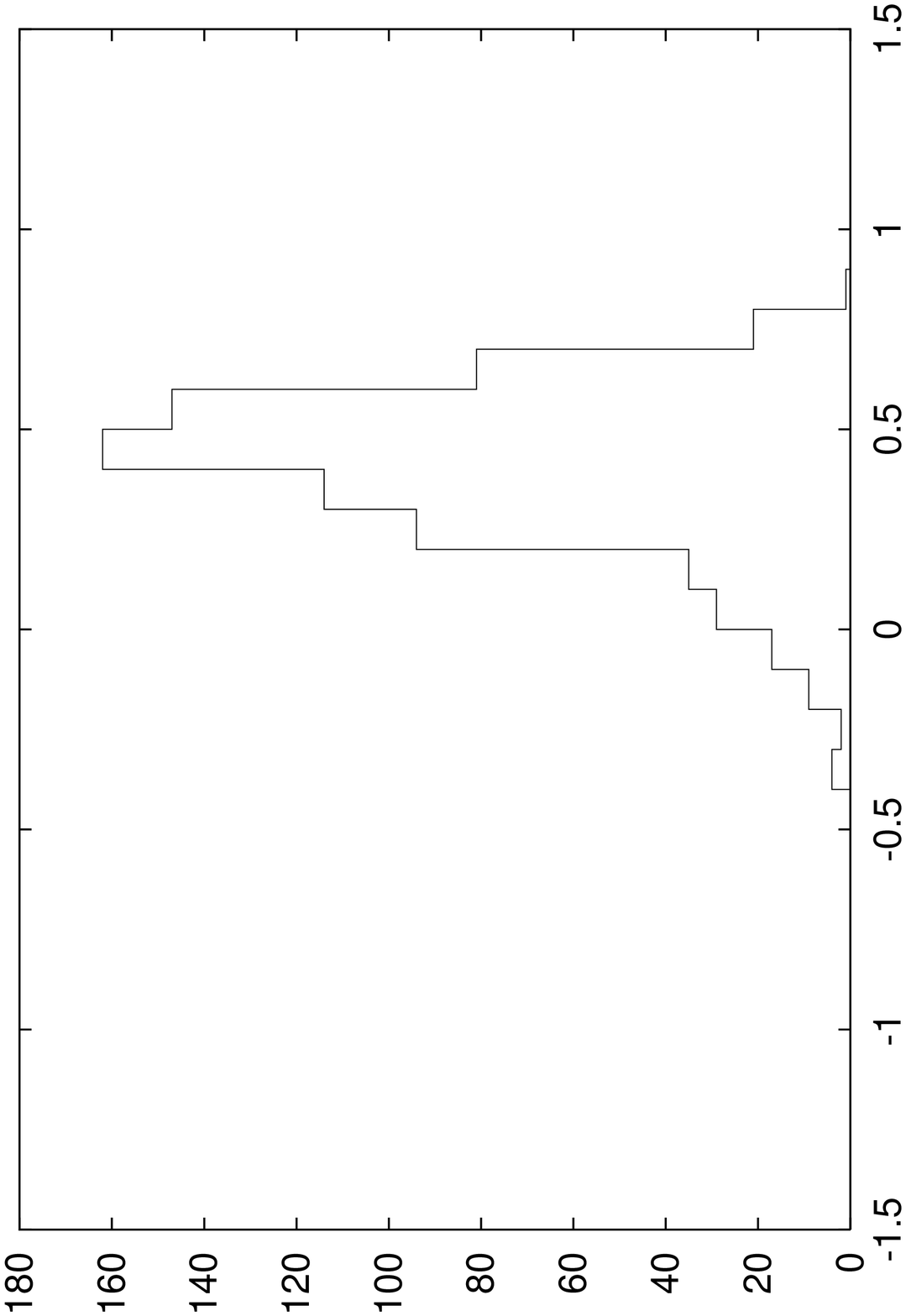}
\vspace{7.5cm}
\includegraphics{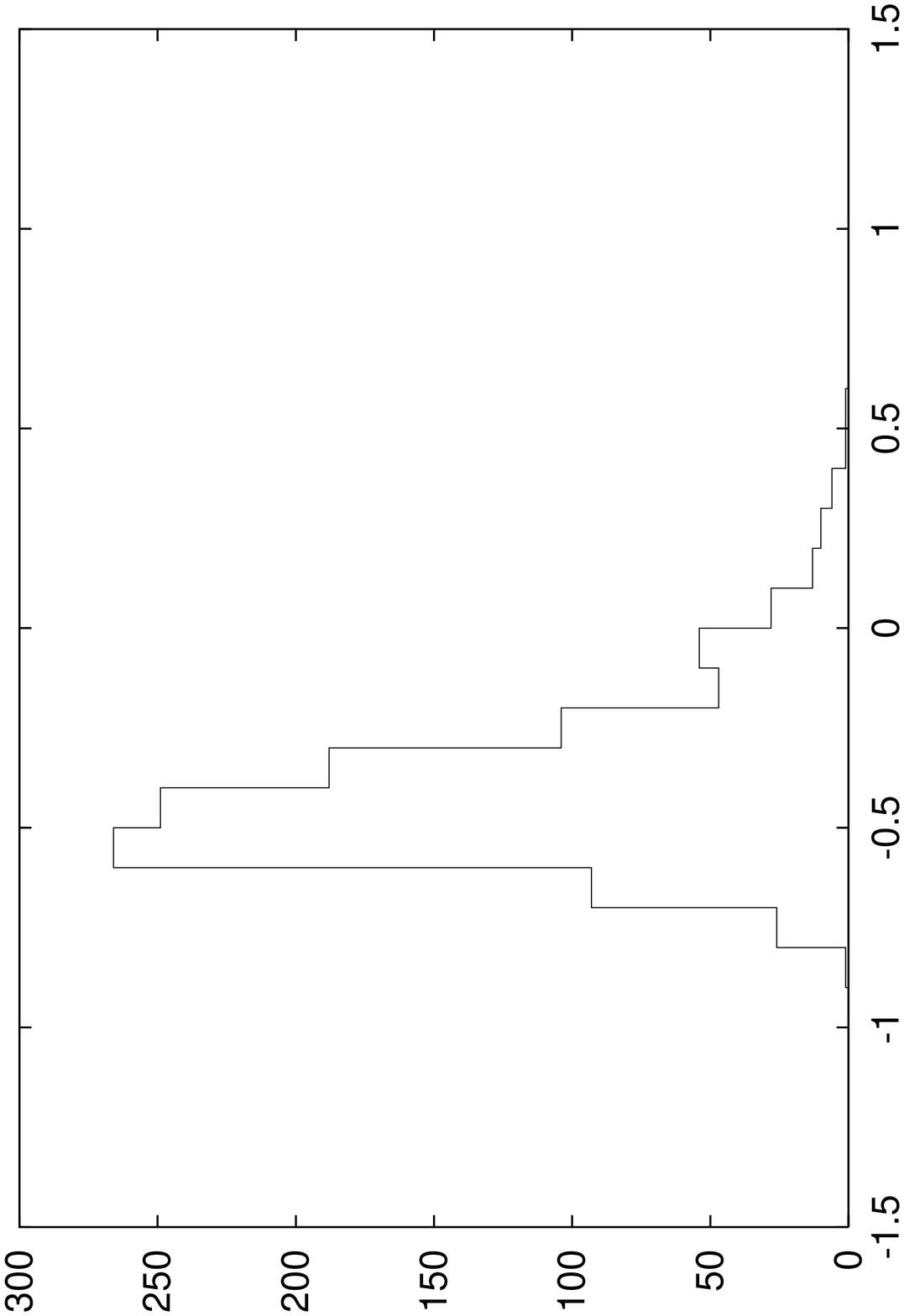}
\vspace{7.5cm}
\includegraphics{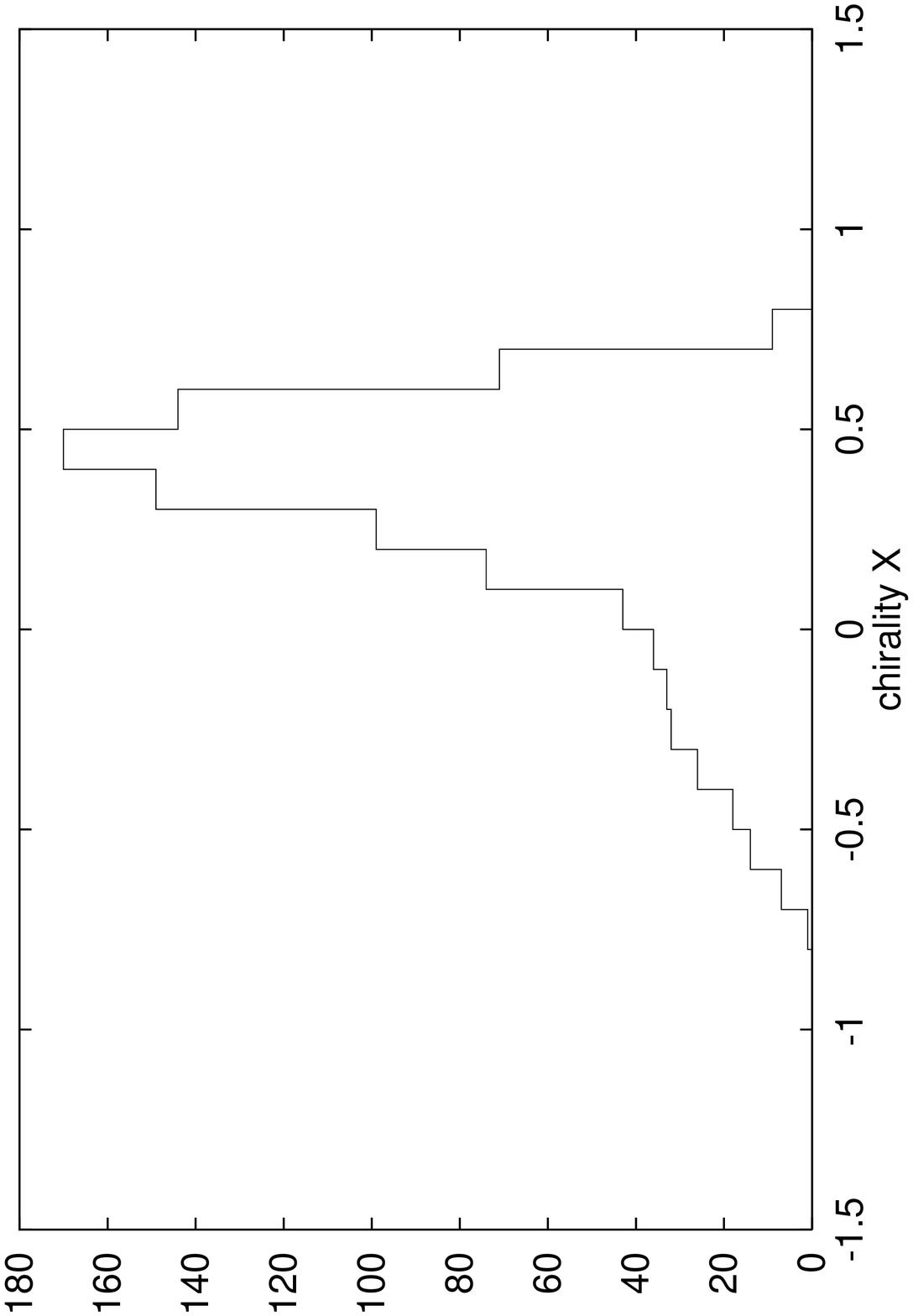}
\vspace{3.0cm}
\caption[]{Chirality histograms for three typical single eigenmodes with exactly real
eigenvalues. Note that these modes are {\it globally} chiral, i.e. they have
either a positive or a negative peak but not both.}
\label{fig:q5mqa}
\end{figure}

\newpage

\vspace*{1.0cm}
\begin{figure}
\includegraphics{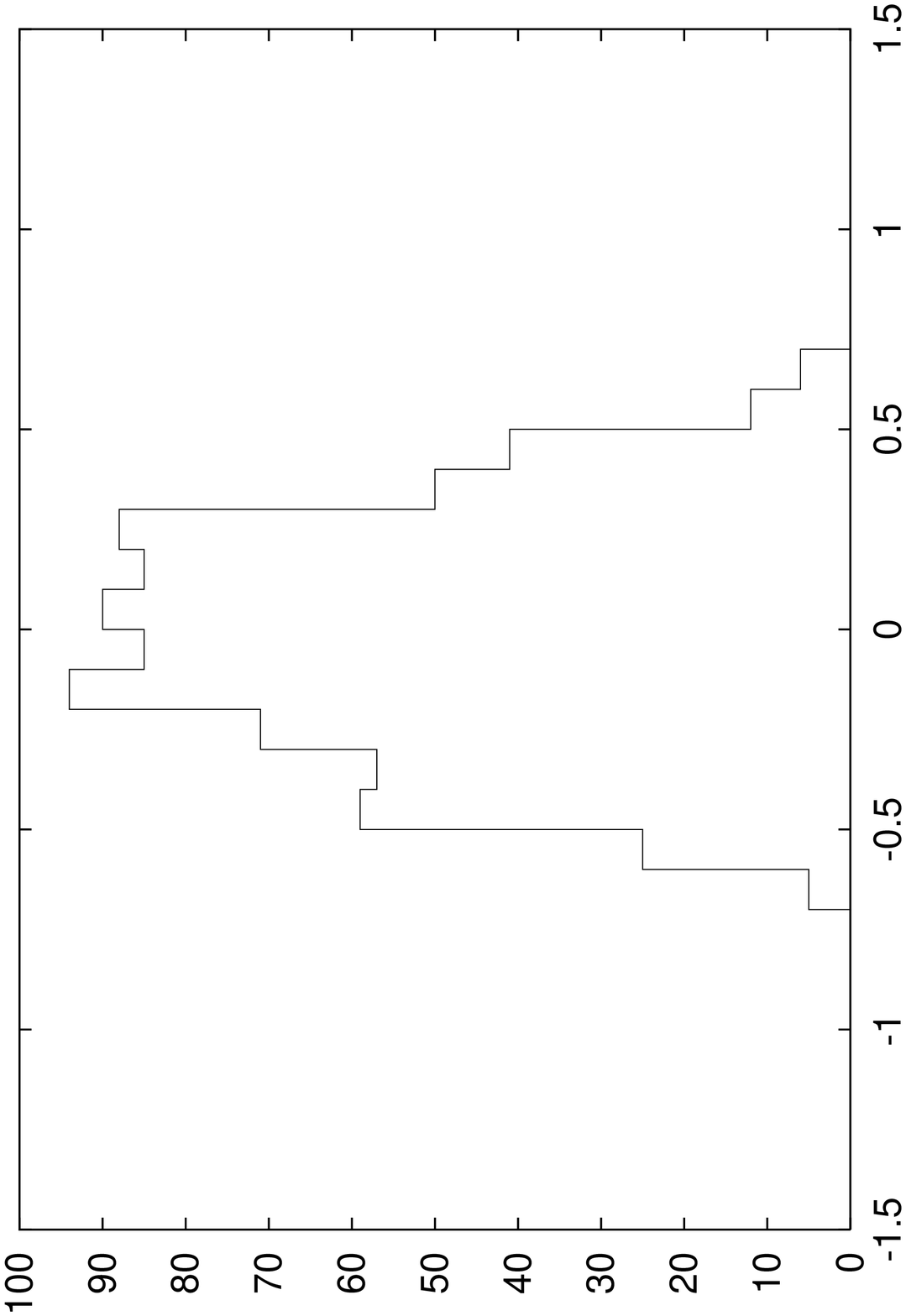}
\vspace{7.5cm}
\includegraphics{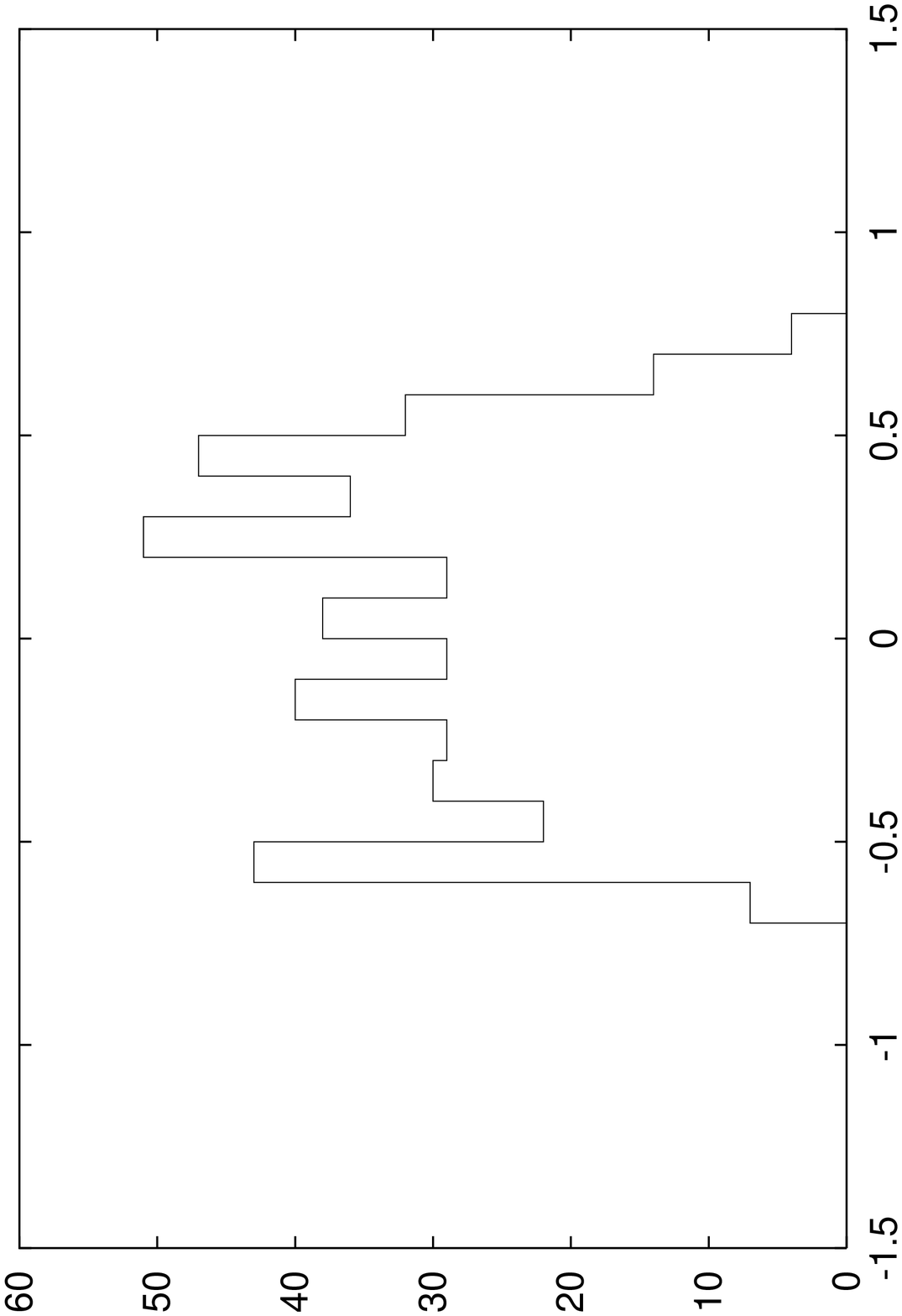}
\vspace{7.5cm}
\includegraphics{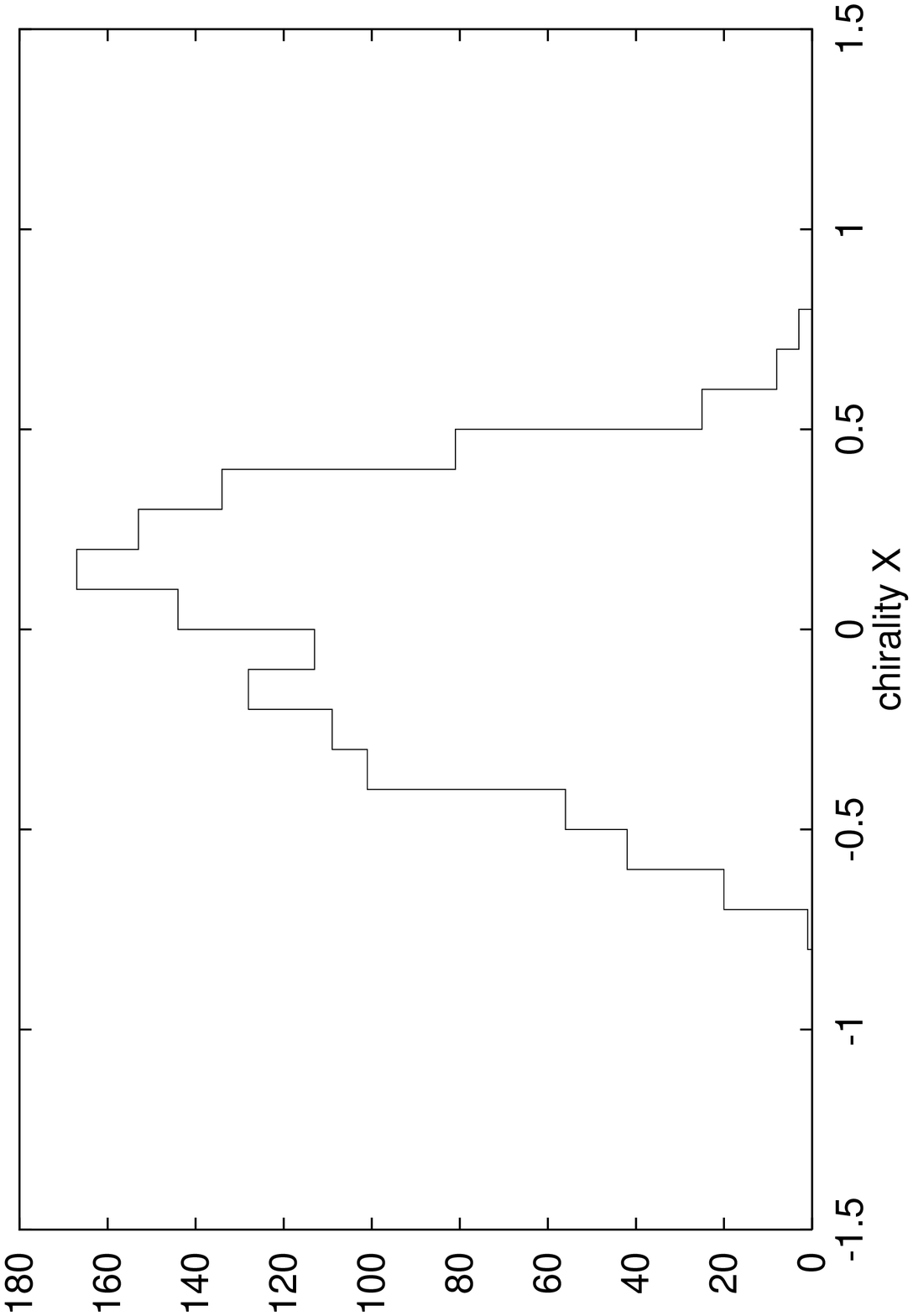}
\vspace{3.0cm}
\caption[]{Chirality histograms for three typical single eigenmodes with complex
eigenvalues very close to, but not on, the real axis. The values of $|{\rm Im}\lambda|$ for these
modes are $.00431a^{-1}, .00828a^{-1}$, and $.0135a^{-1}$, respectively, or about 5 MeV, 10 MeV, and
16 MeV in physical units.}
\label{fig:q5mqa2}
\end{figure}

\vfill\eject

\section {Conclusions}
\label{sec:conclude}

It is now widely accepted that topological charge plays an important role in
low energy QCD. This was originally discovered through the realization that vacuum 
topological transitions can lead to a resolution of the $U_A(1)$ problem.
However, since Witten's 1979 papers, it has been unclear whether the {\it dynamics}
underlying the $\eta'$ mass is associated with the semiclassical tunnelling
events called instantons or with the large vacuum fluctuations typical
of confinement. More generally, it has been unclear whether fluctuations
of topological charge in the QCD vacuum are predominantly in the form of instantons.

In this paper we have  approached this question from the fermionic point 
of view. After reviewing the role of fermionic near-zero modes in $s\chi SB$ and in 
the generation of the $\eta'$ mass, we have argued that the presence or absence of 
instantons is conveniently encoded in the {\it local chirality} properties of these 
near-zero modes. The main points of this paper are (1) to propose the local chirality 
calculation as a way of testing the instanton picture in the framework of lattice
QCD, and (2) to present our lattice data, which convincingly indicate that the fluctuations
of the QCD vacuum are not instanton-dominated, in agreement with Witten's conjecture.  

While our work uses the Wilson-Dirac operator which lacks exact lattice chiral 
symmetry, we are reasonably confident (as we argued extensively in the text)
that our conclusions are not invalidated by lattice artifacts. In this respect, 
we would also like to point out that the connection between (anti)self-duality 
and local chiral properties of low-lying Dirac eigenmodes is particularly 
transparent in QED2. We have tested this on the lattice using the Wilson-Dirac 
operator and the expected chiral structure is indeed exposed even without exact 
lattice chiral symmetry (see the Appendix). Nevertheless, it would 
obviously be quite interesting to see our calculations repeated using a Ginsparg-Wilson 
fermionic operator. While this is more computationally demanding due to 
non-ultralocality~\cite{nonultr}, a study analogous to ours is actually quite 
feasible~\cite{DeGrand-disc}. 

The apparent validity of Witten's conjecture has broad implications for hadron 
phenomenology. Some of these implications are explored in a companion paper
by one of us~\cite{SCQM}. An ongoing effort to construct a data base of low eigenmodes 
for a much larger ensemble of gauge configurations is currently underway. Further 
studies of the chiral and space-time structure of low Dirac eigenmodes should enable
us to extract the detailed structure of topological charge fluctuations in typical
gauge configurations, and hopefully lead to satisfactory phenomenological models.
This should also provide us with a much more precise understanding of chiral symmetry 
breaking and the $U_A(1)$ problem, and allow us to begin to explore the dynamics of 
phenomenologically critical 
quark-pair creation processes, both in hairpin diagrams and in ordinary OZI-allowed 
decays.

\bigskip\bigskip\bigskip

{\centerline {\bf ACKNOWLEDGEMENTS}}

\medskip

The work of N.I. was supported by DOE contract DE-AC05-84ER40150 
under which the Southeastern Universities Research Association (SURA) operates the Thomas 
Jefferson National Accelerator Facility (Jefferson Lab). 
The work of Ivan Horv\'ath, John McCune and 
H.B.~Thacker was supported in part by the Department of Energy under grant DE-FG02-97ER41027.
I.H.~was also supported by DOE contract DE-FG05-84ER40154.

\newpage
\appendix

\section{Local Chirality in QED2}
\label{app:qed2}

To further examine the suitability of local chirality properties of 
Dirac near-zero modes as a test of local self-duality for the underlying 
gauge fields, we studied this connection also in the case of QED2.
This is an ideal testing ground for the underlying ideas we introduced in 
Section~\ref{sec:locchiral} because, in some sense, the gauge field is 
automatically (anti)self-dual. This can be seen from the fact that 
$F_{\mu\nu}\equiv \partial_{\mu}A_{\nu} - \partial_{\nu}A_{\mu}$ 
only has one independent component (the electric field $E=F_{12}$), 
yielding for the Euclidean action density
${\cal L}\propto F_{\mu\nu}F_{\mu\nu} \propto E^2$, 
and for the topological charge density 
$Q(x) \propto \epsilon_{\mu\nu}F_{\mu\nu}\propto E$. Thus, the magnitude
of $Q(x)$ is always locked in with the magnitude of ${\cal L}$, as is the
case for the (anti)self-dual field in four dimensions. We would like 
to stress, though, that in case of QED2 this has nothing to do with 
relevance of instantons, but rather it is a generic property of the
gauge field. 

In terms of the Dirac eigenvalue problem (analogous to 
Eq.~(\ref{eq:schrodleft})-(\ref{eq:schrodright})), one finds that the left and right 
components of the eigenvector $\psi$ in
${\mbox{$i\,\!\!\not\!\!D$}} \psi \,=\, \lambda\psi$ have to satisfy
\begin{eqnarray}
\left[-D^2 \,+\, e E \,\right]\psi_L \,=\, \lambda^2\psi_L \\
\left[-D^2 \,-\, e E \,\right]\psi_R \,=\, \lambda^2\psi_R
\label{eq:schrod2}
\end{eqnarray}
where $D_{\mu}\equiv \partial_\mu + ieA_{\mu}$, and we have used the 
convention $\gamma_5\equiv i\gamma_1\gamma_2$. From this one can
see that in the region around the local positive maximum of $E$ it is 
energetically favorable for the near-zero mode to have a large $\psi_R$
and relatively small $\psi_L$. Similarly, around local negative maxima of 
$E$, one expects the dominance of $\psi_L$. In other words, there should
be a certain degree of local chirality exhibited by the near-zero modes.
(Again, we emphasize that in QED2, unlike QCD4, the local chirality of the
near-zero modes is expected in the vicinity of any strong fluctuation of the gauge field,
and thus does {\it not} imply the existence instantons.)

We have tested this scenario on the lattice using a procedure analogous to the 
one we have discussed at length for QCD. In particular, we have 
calculated the eigenmodes of a Wilson-Dirac operator in the background
of Monte Carlo generated configurations of pure gauge compact QED2,
and calculated local chirality in the regions where the low-lying eigenfunction
is large. In Figures 6 and 7 we show the results from $100$ 
configurations on a $24\times 24$ lattice at $\beta=2$ (we use the standard 
normalization for Wilson's plaquette action). The two histograms show the
local chirality of real modes and of near-real modes lying in the vicinity 
of the continuum branch of the Wilson-Dirac spectrum, respectively. 
In Figure 7 near-real modes with imaginary part less than $0.3$ in 
absolute value were included, while the maximal imaginary part for the high 
modes is typically around $1.3$ (we use the standard normalization for 
the Wilson-Dirac operator). In both cases, about $1\% $ of the highest 
points in a given mode were used. Similar to the situation in QCD, significant 
local chirality is observed for the real modes as expected from the index 
theorem in the continuum. However, contrary to the situation in QCD, chiral 
peaks are also apparent for the non-zero modes in accordance with the above 
arguments. It should be emphasized here that these results not only 
illustrate the validity of the general approach proposed in this paper,
but also further confirm that the required effects can be captured
by the Wilson-Dirac operator. Indeed, we can observe the local chirality
of near-zero modes even though the gauge fields are rather rough, 
even though the cut for imaginary part of the included modes is rather high,
and even though the lumpiness of the topological charge is not expected to be
as pronounced here as in QCD.

\newpage

\vspace{1.0in} 
\begin{figure}\vspace*{2.0cm}
\includegraphics{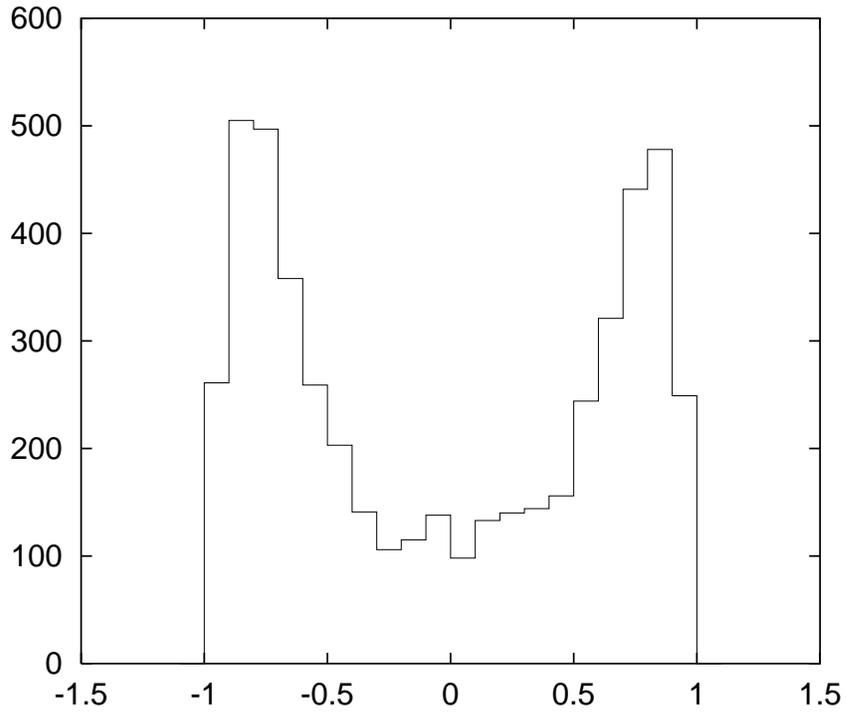}
\vspace{6.0cm}
\caption[]{Chirality histogram for real modes in QED2.  }
\label{fig:real_modes_qed2}
\end{figure}

\vspace{1.0in}
\begin{figure}\vspace*{2.0cm}
\includegraphics{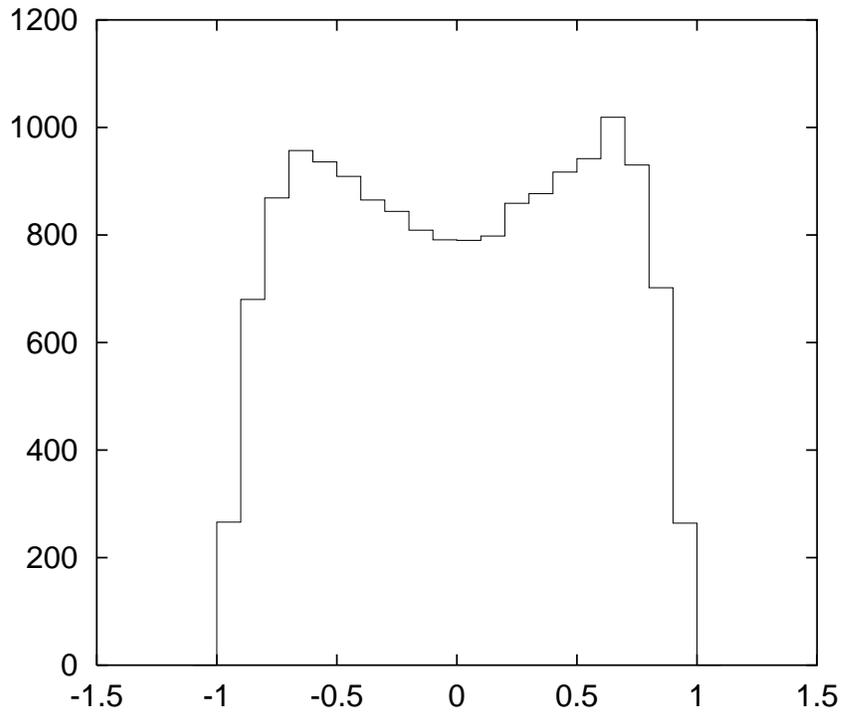}
\vspace{6.0cm}
\caption[]{Chirality histogram for near-real modes in QED2 with complex eigenvalues
$|{\rm Im}\lambda < 0.3|$.}
\label{fig:complex_modes_qed2}
\end{figure}

\vfill
\eject

{

\end{document}